\documentstyle[aps,prd]{revtex}
\begin{document}

\newcommand{\be}{\begin{equation}}
\newcommand{\ee}{\end{equation}}
\newcommand{\ben}{\begin{eqnarray}}
\newcommand{\een}{\end{eqnarray}}
\newcommand{\nn}{\nonumber}
\newcommand{\n}{\label}

\title
{Global monopoles and scalar fields as the electrogravity dual of
Schwarzschild spacetime}

\author{
Naresh Dadhich\footnote{Electronic mail address: nkd@iucaa.ernet.in}} 
\address{Inter University Centre for Astronomy and Astrophysics, Post
Bag 4, Ganeshkhind, Pune 411007, India}
\author{
Narayan Banerjee\footnote{Electronic mail address:narayan@juphys.ernet.in}} 
\address {Relativity and Cosmology
Research Centre, Department of Physics, Jadavpur University, 
Calcutta-700032, India}

\date{\today}

\maketitle

\begin{abstract}
We prove that both global monopole and minimally coupled static zero mass 
scalar field are electrogravity dual of the Schwarzschild solution or flat 
space and they share the same equation of state, $T^0_0 - T^i_i = 0$. This 
property was however known for the global monopole spacetime while it is for 
the first time being established for the scalar field. In particular, it turns
out that the Xanthopoulos - Zannias scalar field solution is dual to flat 
space. 
\end{abstract} 

\hspace{5mm}

Phase transitions in the early universe might have given rise to
several kinds of topological defects depending on the nature of the
symmetry that is broken\cite{kibble}. If a global SO(3) symmetry of a
triplet scalar field is broken, the point like defects called global
monopoles are believed to be formed. Barriola and
Vilenkin\cite{barriola} presented a solution which describes a global
monopole at a large radial distance. It gives back the usual Schwarzschild 
spacetime when the monopole charge is put equal to zero.   \\  \\
Very recently it has been shown that in terms of the electrogravity
duality, the dual to the static spherically symmetric vacuum
solutions are the global monopole solutions\cite{dadhich1}. Like the
Maxwell field, the gravitational field can also be resolved into two
parts, the electric part, generated by the source distribution and
the magnetic part, brought into being by the motion of the 
sources\cite{lanczos}. The electric part can be further decomposed
into two parts corresponding to two kinds of sources; non-gravitational energy
distribution giving rise to active part and gravitational field energy to 
passive part. A duality transformation
can be defined between them which leaves the Einstein--Hilbert action
invariant if the Newtonian constant of gravitation $G$ is replaced
by $-G$ \cite{dadhich2}. Electrogravity duality, in particular, is defined by
interchange of active and passive electric parts which in familiar terms 
translates into interchange of the Ricci and Einstein tensors. This dulaity
 is at the more fundamental level the interchange between the Riemann 
curvature and its double dual (left and right dual). Note that contraction of 
Riemann is Ricci while its double dual is Einstein tensor. This is why it 
implies interchange of Ricci and Einstein tensors. Though the 
Einstein vacuum equation is invariant under this duality, it is however 
possible to 
find dual solution to the Schwarzschild solution. This happens because 
the effective vacuum equation can be given which is duality non-invariant 
and yet giving the same Schwarzschild solution \cite{dadhich1}. The Barriola 
- Vilenkin global monopole solution [2] turns out to be dual to the 
Schwarzschild solution.  \\  \\
The remarkable thing is that in a similar manner it turns out  
that the minimally coupled static massless scalar field solution is also dual
to the Schwarzschild solution. This happens perhaps because both the 
constructs are different 
manifestations of the scalar field configuration and they satisfy the same 
equation of state. This means that there is yet another independent way of 
getting to the 
Schwarzschild solution which leads to a different dual solution than that of 
the global monopole. Further both global monopole and scalar field also exist 
in the forms which are dual to flat space. The dual flat solutions are 
contained in the dual Schwarzschild solutions and could be obtained by 
putting the appropriate parameter to zero. Global monopole spacetime is  
asymptotically non-flat while the scalar field spacetime is on the other hand 
asymptotically flat. \\ \\ 
Although the global monopole solution was known 
to be dual of the Schwarzschild solution [3], that the scalar field also 
shares this property is being established for the first time. It is quite 
remarkable and interesting. For a priori there 
is nothing much common between the two except being different manifestations 
of scalar field configuration. Both could in a sense be considered as 
modification of the Schwarzschild/flat space as they are in the Newtonian 
limit indistinguishable from the Schwarzschild/flat space. The one produces 
asymptotically non-flat while the other asymptotically flat 
modification. \\ \\ 
Let us begin by defining the three kinds of energy density as follows:
(i) Energy density as measured by a static observer, $\rho =
T_{ab}u^au^b;    
u_au^a = 1$. (ii) Null energy density, $\rho_n = T_{ab}k^ak^b;   k_ak^a
= 0$.  (iii) Timelike convergence density, $\rho_t = (T_{ab} - 1/2~Tg_{ab})
u^au^b$. We shall use them to characterize effective vacuum and its dual 
space. It turns out that vacuum for spherically symmetric spacetime could 
effectively be characterized by the condition, $\rho = 0, \rho_n = 0$. 
This definition is physically illuminating as it is parallel to the Newtonian 
definition which simply demands vanishing of matter density\cite{dadhich3}. 
Since in GR, we want both timelike and 
null particles to interact gravitationally, hence density relative to both 
of them must vanish in vacuum. $\rho = 0$ would imply $G^0_0 = 0$ while 
$\rho_n = 0$ would imply either $G^0_0 = G^1_1$ for radial or 
$G^0_0 = G^2_2$ for circular or $G^0_0 = G^1_1 = G^2_2$ for non radial 
photons. Of course the last alternative with $\rho = 0$ would imply the 
usual vacuum equation $R_{ab} = 0$. However either of the other two with 
vanishing $\rho$ is good enough to give the Schwarzschild solution. We thus 
have the effective empty space equation as
\be
G^0_0 = G^1_1 = 0,
\ee
or alternatively
\be
G^0_0 = G^2_2 = 0.
\ee
Note that for the spherical symmetry, the vacuum equation implies only two 
independent equations, which would not in contrast to $R_{ab}=0$ be in general
electrogravity duality ($R_{ab}\leftrightarrow G_{ab}$) invariant. Thus under 
the duality transformation, the above equations would give rise to the 
two sets of the dual equations which would admit two distinct dual solutions. 
\\ \\
Of course both of the above equations admit the well-known Schwarzschild 
solution as the general solution, which is given by the metric,
\be
ds^2 = Adt^2 - A^{-1}dr^2 -r^2d\omega^2, ~d\omega^2 = d\theta^2 + 
sin^2\theta d\varphi^2,
\ee
where
\be
A = 1 -2m/r,
\ee
with $m$ denoting the mass of the Schwarzschild particle. \\ \\
The electrogravity duality is defined\cite{dadhich1} by the interchange of 
active and passive electric parts which translates in the familiar terms to the
interchange of Ricci and Einstein tensors, viz.
\be
R_{ab}\leftrightarrow G_{ab}.
\ee \\ \\
The equation dual to the effective vacuum equations would be
\be
R^0_0 = R^1_1 = 0,
\ee
corresponding to eqn(1) and
\be
R^0_0 = R^2_2 = 0.
\ee 
corresponding to eqn(2). For densities, the duality transformation implies 
$\rho\leftrightarrow\rho_t, \rho_n\rightarrow\rho_n$. \\ \\
The former set (6) admits the general solution as the Barriola - Vilenkin 
global monopole solution\cite{barriola} given by
\be
ds^2 = (A -\eta^2)dt^2 -(A -\eta^2)^{-1}dr^2 -r^2d\omega^2,
\ee
where $\eta$ marks the scale of symmetry breaking. The stresses are given by
\be
T^0_0 = T^1_1 = \eta^2/r^2,
\ee
which is the form required at large distance by global monopole\cite{barriola}
. Note that it is not flat even when $m=0$ and hence is asymptotically 
non-flat. In this limit it turns out to be dual to flat space. \\ \\ 
The latter set (7) would admit the general solution given by
\be
ds^2 = A^{n}dt^2 -A^{-n}dr^2 -r^2A^{1-n}d\omega^2,
\ee
where $A$ is as given in eqn (4). Interestingly this is also dual to the 
Schwarzschild solution to which it would reduce when 
$n=\pm1$. It reduces to flat space when $m=0$ and hence is clearly 
asymptotically flat. \\ \\
The stress tensor for a scalar field reads as
\be
T_{ab} = \phi_{,a}\phi_{,b} -\frac{1}{2}\phi_{,c}\phi^{,c}g_{ab},
\ee
which for the static spherically symmetric field would have $T^0_0 = -T^1_1 
= T^2_2$. For the above metric, we shall thus have 
\be
T^0_0 = -T^1_1 = T^2_2 = (1-n^2)\frac{m^{2}}{r^{4} A^{(2-n)}},
\ee
and the scalar field itself is given by
\be
\phi =(1-2m/r)^{\sqrt{\frac{1-n^2}{2}}}.
\ee
Clearly when $n=1$, it reduces to the Schwarzschild solution while for $n=-1$ 
it can be brought to the standard Schwarzschild form by 
letting $m\rightarrow-m$. Note that for acceleration to be
attractive for the metric (10, $m$ must be positive while the scalar field 
energy density in eqn (12) is 
insensitive to its sign. The above solution thus reduces to the vacuum 
spacetime which is respectively the Schwarzschild ($m$)and anti-Schwarzschild 
($-m$) according to $n=1$ and $n=-1$. \\ \\ 
On the other hand the case $n=0$ would be the case of dual flat (dual set (7) 
for the gauge $g_{00}=1$) scalar field solution. It could be cast into the 
form, 
\be
ds^2 = dt^2 -dr^2 -(r^2-m^2)d\omega^2.
\ee
and the stress tensor would reduce to 
\be
G^0_0 = - G^1_1 = G^2_2 = -\frac{m^2}{(r^2-m^2)^2},
\ee
with the scalar field,
\be
\phi = \frac{1}{\sqrt2}ln\frac{r-m}{r+m}.
\ee 
It is the Xanthopoulos and Zannias (XZ) scalar field solution
\cite{xz} which is dual to flat space.  \\ \\
We have thus two independent two parameter families of solutions dual to the 
Schwarzschild solution which contain the dual flat families. One of them is 
asymptotically flat while the other is not. Both share the same equation of 
state, $\rho_t=0$. \\ \\
To the scalar field a global monopole could readily be added by a general 
prescription due to Dadhich and Patel\cite{dlk} for any spherically 
symmetric solution. We
just need to multiply the angular part of the metric by a constant. Thus in the
above XZ scalar field metric (14) a global monopole could be added simply by 
writing $k^2(r^2-m^2)$ in place of $r^2-m^2$. It would have the 
superposition of the two stress tensors; $T_{ab} = T_{ab}(GM) + T_{ab}(SF)$. 
For the global monopole we have,
\be
T^0_0 = T^1_1 = \frac{\left( 1 - 1/k^2 \right)}{r^2-m^2}.   
\ee
and for the XZ scalar field
\be
T^0_0 = -T^1_1 = T^2_2 = \frac{m^2}{(r^2-m^2)^2}.
\ee
It thus represents a massless global monopole superposed onto the XZ scalar 
field\cite{bd}. It reduces to the massless global monopole (8) for $m=0$ and
to the XZ scalar field (14) for $k=1$. \\ \\
Further, it is similarly also possible to superpose the other two dual 
solutions of global monopole (8) and of the scalar field (10). The resulting 
metric would read as
\be
ds^2 = (A -\eta^2)^{n} -(A -\eta^2)^{-n}dr^2 -r^2(A -\eta^2)^{1-n}d\omega^2,
\ee
which is identified to be the global monopole  with a scalar
field as given by Banerjee et al\cite{banerjee}.
Note that the above metric could be transformed to the form in which the 
angular part of the scalar field metric is multiplied by a constant. The 
scalar field $\phi$ as given by (13) and the equation of state $\rho_t = 0$ 
would remain unaltered under this superposition. The stress tensor for the 
metric (19) would be given by 
\ben
T^0_0 &=& \frac{\eta^2/r^2}{(A - \eta^2)^{1 - n}} + T_2^2 \\
T^1_1 &=& T^0_0 - 2 T^2_2 \\
T^2_2 &=& (1 - n^2) \frac{m^2/r^4}{(A - \eta^2)^{(2-n)}}
\een
which is the superposition of the global monopole and the scalar field. 
Clearly it is the global monopole for $m=0$ and the scalar field for 
$\eta = 0$. \\ \\
By utilising the fact that the Schwarzschild solution could be obtained by two 
distinct sets of equations which are electrogravity duality non-invariant, we 
have shown that both the global monopole and the scalar field solutions are 
dual to the Schwazschild solution. They also include solutions dual to flat 
space. The other interesting feature is that the global 
monopole spacetimes are asymptotically non-flat while those of the scalar 
field are asymptotically flat. It is also remarkable that the global monopole 
and the scalar field could be 
superposed in which the resulting stress tensor is the sum of the two, and 
the only change in the scalar field metric is a constant multiplying the 
angular part. This is in accordance with the general prescription of Dadhich 
and Patel\cite{dlk} for intrduction of a global monopole like stresses in any spherically symmetric spacetime. All the spacetimes discussed here satisfy 
the same equation of state $\rho_t=0$, including the one in which the two 
fields are superposed. Under the superposition the solution for the scalar 
field remains undisturbed.  \\ \\
 From the stresses in eqns (12) and (15), it is clear that the scalar field 
has energy density which falls of as $1/r^4$, like the gravitational field 
energy density, which could be made both positive or negative by choosing the 
sign of constant $m^2$. It is however generally taken to be positive as is 
the case for the XZ solution. Thus the scalar field solution 
represents a Schwarzschild particle sitting in a positive energy distribution. 
Very recently 
Dadhich\cite{dadhich4} has shown that for a distribution engulfing an 
isolated object, the ``positive'' energy condition is that its energy is 
negative. Its effect like that of gravitational field energy would be felt 
only through the space curvature. Since gravitational field energy is negative
, it produces negative curvature in space which works in unison with the 
attraction produced by the gradient of potential. Thus only when energy 
density is negative, it would act in conformity with the 
acceleration produced by the gradient of potential. Since the scalar field 
energy density is positive it would produce positive  
space curvature which would tend to oppose the effect of the potential. This 
is similar to the case of charged black hole where positive electric field 
energy density produces the repulsive effect counteracting the effect of 
mass. A 
similar situation also obtains for a black hole on the brane in the currently 
popular Randall-Sundrum brane world model\cite{rs}. The back reaction of the 
bulk on the brane effectively produces on the brane a trace free matter field 
through the Weyl curvature of the bulk. That is the black hole on the brane 
has the similar enviornmemnt\cite{dadhich etal} of energy density falling off 
as $1/r^4$. Here also if we let energy density of scalar field to be 
negative, it would 
then contribute positively to the field of the Schwazschild particle as in 
the brane world model. 
That is we can approximately mimick the black hole on the brane  simply 
by a negative energy density scalar field. This is an interesting and novel 
way of looking at the scalar field solution. We hope to work further on this 
track in future.  \\ \\
The XZ scalar field solution (14) could be cast in the following interesting 
form,
\be
ds^2 = dt^2 -(1 + m^2/r^2)^{-1}dr^2 -r^2d\omega^2,
\ee
which clearly shows its asymptotic flatness. There is no central mass to 
produce the radial acceleration. Hence the motion for both the dual flat 
global monopole (metric (8) with $m=0$) as well as the dual flat XZ scalar 
field solutions would be solely 
guided by the space curvature. So long as the energy density is positive, 
there would occur repulsive effect [11] as is well-known for the global 
monopole\cite{hl}. The contrary would be the case for negative energy density. 
\\ \\
Finally it is worth remembering that the duality character of the scalar 
field solutions unlike that of global monopole is being establishsed for the 
first time. It would be interesting to seek scalar field dual solutions in 
all those cases where global monopole solutions exist, for instance in the 
NUT space\cite{ndl}, Kaluza - Klein space\cite{dlt} and 2+1 gravity\cite
{bdk}.\\ \\

Acknowledgement: NB wishes to thank IUCAA, where the major part of this work 
was done for warm hospitality. \\


\begin{thebibliography}{100}
\bibitem{kibble}
T.W.B.Kibble, J.Phys.A: Math. Gen, {\bf 9}, 1387 (1976); A.Vilenkin, Phys.Rep.,
{\bf 121}, 263 (1985).
\bibitem{barriola}
M.Barriola and A.Vilenkin, Phys. Rev. Lett. {\bf 63}, 341 {1989}.
\bibitem{dadhich1}
N.Dadhich, Mod. Phys. Lett.A, {\bf 14}, 337 (1999) ; N. Dadhich, Gen. Relativ.
Grav. {\bf32}, 1009 (2000).
\bibitem{lanczos}
C.Lanczos, Ann. Math., {\bf 39}, 842 (1938); L.Bell, C.R.Acad.Sci.,
{\bf 248}, 1297 (1959); M.A.G.Bonilla and J.M.M.Senovilla, Gen.
Relativ. Gravit. {\bf 29}, 91 (1997).
\bibitem{dadhich2}
N.Dadhich, Mod. Phys. Lett. A, {\bf 14}, 759 (1999).
\bibitem{dadhich3} 
N. Dadhich, Current Science {\bf 78}, 1118 (2000).
\bibitem{xz} 
BC Xanthopoulos and T. Zannias, Phys. Rev. {\bf D40}, 2564 (1989).
\bibitem{dlk} 
N. Dadhich and L.K. Patel, Pramana, {\bf 52}, 359 (1999).
\bibitem{bd} 
S. Bose and N. Dadhich, Phys. Lett. {\bf B488}, 1 (2000).
\bibitem{banerjee}
A.Banerjee, A.Beesham, S.Chatterjee and A.A.Sen, Class. Quantum Grav.,
{\bf 15}, 645 (1998).
\bibitem{dadhich4} 
N. Dadhich, Phys. Lett. {\bf B492}, 357 (2000) .
\bibitem{rs} 
L. Randall and R. Sundrum, Phys. Rev. Lett. {\bf 83}, 3370 \& 4690 (1999).
\bibitem{dadhich etal}
N. Dadhich, R. Maartens, P. Papadopoulos and V. Rezania, Phys. Lett. {\bf B487}, 1 (2000).
\bibitem{hl}
D. Harari and C. Lousto, Phys. Rev. {\bf D42}, 2626 (1990); N. Dadhich, K.
Narayan and U. Yajnik, Pramana, {\bf 50}, 307 (1998).
\bibitem{ndl} 
M. Nouri-Zonoz, N. Dadhich and D. Lynden-Bell, Class. Quant. Grav. {\bf 16}, 1021 (1999).
\bibitem{dlt} 
N. Dadhich, L.K. Patel and R. Tikekar, Mod. Phys. Lett. {\bf A14}, 2721 (1999).
\bibitem{bdk} 
S. Bose, N. Dadhich and S. Kar, Phys. Lett. {\bf B477}, 451 (2000).
\end{thebibliography}
\end{document}